\def\be{\begin{equation}}
\def\ee{\end{equation}}
\def\ba{\begin{array}}
\def\ea{\end{array}}

\documentclass[preprint,showpacs,amsmath]{revtex4}
\usepackage{amssymb}
\usepackage{amsmath,amstext,amsbsy}
\def\qed{\leavevmode\unskip\penalty9999 \hbox{}\nobreak\hfill
     \quad\hbox{\leavevmode  \hbox to.77778em{%
               \hfil\vrule   \vbox to.675em%
               {\hrule width.6em\vfil\hrule}\vrule\hfil}}
     \par\vskip3pt}

\newtheorem{theorem}{Theorem}

\newtheorem{proposition}[theorem]{Proposition}
\newtheorem{lemma}[theorem]{Lemma}
\begin{document}

\title{SLOCC Invariants for Multipartite Mixed States}
\author{Naihuan Jing$^{1, 5}$}
\email{jing@math.ncsu.edu}
\author{Ming Li$^{2, 3}$}
\author{Xianqing Li-Jost$^{2}$}
\author{Tinggui Zhang$^{2}$}
\author{Shao-Ming Fei$^{2, 4}$}
\affiliation{
$^1$School of Sciences, South China University of Technology, Guangzhou 510640, China\\
$^2$Max-Planck-Institute for Mathematics in the
Sciences, 04103 Leipzig, Germany\\
$^3$Department of Mathematics, China University of Petroleum , Qingdao 266555, China\\
$^4$School of Mathematical Sciences, Capital Normal
University, Beijing 100048, China\\
$^5$Department of Mathematics, North Carolina State
University, Raleigh, NC 27695, USA}

\begin{abstract}
We construct a nontrivial set of invariants
for any multipartite mixed states under the SLOCC symmetry. These
invariants are given by hyperdeterminants and independent
from basis change. In particular, a family of $d^2$ invariants for
arbitrary $d$-dimensional even partite mixed states are explicitly given.
\end{abstract}
\maketitle

\section{Introduction}

Classification of multipartite states
under stochastic local operations and classical communication (SLOCC) has been a central
problem in quantum communication and computation.
Recently advances have been made for the classification
of pure multipartite states under SLOCC \cite{GW, YQS} and
the dimension of the space of homogeneous SLOCC-invariants in a fixed degree is given
as a function of the number of qudits.
In this work we present a general method to construct
polynomial invariants for mixed states under SLOCC.
In particular we also derive general
invariants under local unitary (LU) symmetry for
mixed states.

Polynomial invariants have been investigated in
\cite{Rains, Grassl, Laura, FJ}, which allow in
principle to determine all the invariants of local unitary
transformations. However, in practice it is
a daunting task to solve the large system of
algebraic equations. Moreover,
little is known for mixed multipartite states
even for $2\times 2\times 2$-systems. In
\cite{makhlin} a complete set of 18 polynomial invariants is
presented for the locally unitary equivalence of two qubit-mixed
states.
Partial results have been obtained for three qubits states
\cite{linden}, tripartite pure and mixed states \cite{SCFW}, and
some generic mixed states \cite{SFG, SFW, SFY}. Recently the local
unitary equivalence problem for multiqubit \cite{mqubit} and general
multipartite \cite{B. Liu} pure states have also been solved.

In general there are limited tools available to resolve the SLOCC symmetry for
mixed multipartite states. Ideally it is hoped that a
complete set of invariants under local unitary transformations can be found.
Nevertheless usually these invariants depend on the detailed
expressions of pure state decompositions of a state. For a given
state, such pure state decompositions are infinitely many.
Particularly when the density matrices are degenerate, the problem
becomes more complicated.

In this paper, we present a general method to construct non-trivial invariants under SLOCC
symmetry group. All invariants found using this method are a priori LU-invariants,
as the LU symmetry group is contained in the SLOCC symmetry group.
Therefore SLOCC-invariants are generally much fewer than
LU-invariants.
Very few information is known about the SLOCC-invariants
except for that the degree of the invariant subring \cite{GW},
and even less is known about how to systematically construct
the invariants. We will give a general method to construct
four families of SLOCC-invariants for any mixed state of
even-partite plus an iterative method to construct
general fundamental invariants in all dimensions.

We show that the Cayley's two hyperdeterminants \cite{Ca, gel} can be used
to construct such invariants.
Cayley's second hyperdeterminant has been used to study
the entanglement measure like concurrence
\cite{Hill,wott,uhlm,rungta,ass} and 3-tangles \cite{coff}. It also played
an important role in classification of multipartite pure states
\cite{miy,Luq,vie}. In \cite{ZJLZF} we proved that the hyperdetermiants
give LU-invariants that are independent from the pure state decomposition
and base transition.
One of our current main results is that
the generalized characteristic
polynomials give actually SLOCC invariants and we further show that
in the even partite cases the Cayley's first hyperdeterminant
also lead to some of the nontrivial SLOCC invariants for mixed states.

\section{Hyper-Matrix representations of mixed states}

Let $H$ be the state space of the mixed multipartite quantum state:
$H=H_1\otimes H_2\otimes\dots\otimes H_n$,
where $H_k$ is a Hilbert space of dimension $d_k$ with inner product
$\langle\  ,\  \rangle_i$. Suppose $\sigma^{(k)}_i$ $(i=0, \ldots, d_k^2-1$) are orthonormal basis
of hermitian operators in $End(H_i)$. Here we can take the Gell-Mann basis,
where $\sigma_0^{(i)}=I$ is the identity. In particular, they are Pauli spin
matrices when $d_i=2$.

Then $\rho$ can be expressed as
\begin{equation}
\rho=\sum_{i_1\cdots i_n}a_{i_1\cdots i_n}\sigma_{i_1}^{(1)}\otimes \cdots\otimes\sigma_{i_n}^{(n)},
\end{equation}
where the $n$-dimensional matrix or hypermatrix $A=(a_{i_1\cdots i_n})$ is called the matrix representation
of $\rho$ with respect to the bases $\{\sigma_i^{(k)}\}$. Here the format
of the matrix $A$ is $d_1^2\times \cdots \times d_n^2$, and usual rectangular matrices of size
$m\times n$
are $2$-dimensional matrices of format $m\times n$. One can use $A=[\rho]_{\sigma}$ to denote
the matrix representation of $\rho$ with respect to the orthonormal bases $\sigma=\{\sigma_i^{(k)}\}$.
Sometimes the reference to $\sigma$ is omitted if it is clear from the context.

We recall the multiplication of the hyper-matrix $A$ of format
$f_1\times \cdots \times f_n$ by a $f_i\times f_i$-matrix $B$ as follows:
\begin{equation}
B*_k A=C,
\end{equation}
where
$$
C_{i_1\cdots i_n}=\sum_{j=1}^{f_k} b_{i_kj}a_{i_1\cdots i_{k-1}ji_{k+1}\cdots i_n}.
$$
This generalizes the notion of matrix multiplication. When $A$ is a regular rectangular matrix,
then $B*_1A=BA$, $B*_2A=AB^t$ and $(B*_1C*_2)A=BAC^t$, where $t$ stands for transpose.

It is easy to see that the matrix representation of $\rho$ under basis change from $\sigma$ to $\sigma'$
can be simply described by
\begin{equation}\label{action-P}
[\rho]_{\sigma'}=(P_1*_1\cdots P_n*_n)[\rho]_{\sigma},
\end{equation}
where the unitary matrix $P_k$ is the transition matrix from the orthonormal basis $\{\sigma_i^{(k)}\}$ to
the orthonormal basis $\{\sigma_i'^{(k)}\}$ for $k=1, \cdots, n$.

A function $F(\rho)$ is invariant under SLOCC transformation
if $F(\rho)=F(A\rho A^\dag)$ for any invertible $A=A_1\otimes\cdots\otimes A_n\in G=SL(d_1)\otimes \cdots\otimes SL(d_n)$,
the SLOCC symmetry group, where $\dag$ denote transpose and conjugate.
A true SLOCC-invariant
should only depend on $\rho$, i.e., an invariant $F(\rho)$
given in terms of the matrix $[\rho]$ should be
independent from the matrix representation
\cite{ZJLZF}. Since our matrix representation is given as an un-normalized
form, the SLOCC-invariants can be taken as a relative invariant
under the larger group $\hat{G}=GL(d_1)\otimes\cdots\otimes GL(d_n)$,
or a projective invariant under $\hat{G}$.

Two hypermatrices of the same format can be added together to give rise to a third hypermatrix
by $(a_{i_1\cdots i_n})+(b_{i_1\cdots i_n})=(a_{i_1\cdots i_n}+b_{i_1\cdots i_n})$.
The scalar product of a hypermatrix by a number is defined as usual.
Then the set of hypermatrices of format $f_1\times\cdots\times f_n$ forms
a vector space of dimension $f_1\cdots f_n$.
We denote the vector space by $Mat(f_1, \cdots, f_n)$.

We consider invariants under the SLOCC symmetry.
Let us first look at the special example of bipartite case
to motivate our later discussion.
Suppose $A$ is the (square) matrix representation of
the mixed state $\rho$ on $H^{\otimes 2}$. Using the same method of \cite{ZJLZF}
it is easy to see the following result.

\begin{proposition} Suppose $A$ is the Bloch matrix representation
of the bipartite mixed state $\rho$ on $({\mathbb C}^{d})^{\otimes 2}$. Then the coefficients $F_i(A)$,
$i=1,2,..., d^2$, of the characteristic polynomials of $A$:
\begin{align}\nonumber
\det(\lambda\,I-A)&=\lambda^d-Tr(A)\lambda^{d-1}+\cdots+(-1)^ddet(A)\\
&=\sum_{i=0}^{d}\lambda^{d-i}F_{i}(A) \label{thm}
\end{align}
are LU-invariants. In particular, $tr(A)$ and $\det(A)$
are LU-invariants of the two-partite state $\rho$.
\end{proposition}

We remark that when $\rho$ is a bipartite mixed state in arbitrary dimensional space
$H_1\otimes H_2$, then the Bloch matrix representation $A$ is a rectangular
matrix. Then $det(AA^T)$ and $det(A^TA)$ are LU-invariants. When $m\neq n$,
it seems that these are the only known SLOCC-invariants
according to \cite{GW}.

For pure states $|\phi\rangle=\sum_{i_1, \cdots, i_n}a_{i_1\cdots i_n}|i_1\cdots i_n\rangle$, we can
associate the hypermatrix $A(|\phi\rangle)=(a_{i_1\cdots i_n})$, then the
format will be simpler. For bipartite pure state $|\phi\rangle$, the determinant $det(|\phi\rangle)$
is clearly a SLOCC-invariant.

In the following we extend the result in Proposition 1 to show that $F_i(A)$
are in fact SLOCC-invariants.

\section{Hyperdeterminants and SLOCC-invariants}

Given any hypermatrix $A$, one defines the associated
multilinear form $f_A: 
H_1\otimes \cdots
\otimes H_n\mapsto \mathbb C$ given by
\begin{equation}\label{eq:form}
f_A(x^{(1)}, \cdots, x^{(n)})=\sum_{i_1,\cdots, i_n}a_{i_1\cdots i_n}x_{i_1}^{(1)}\cdots x_{i_n}^{(n)}.
\end{equation}
The multilinear form $f$ can also be written as a
tensor in $H_1^*\otimes\cdots \otimes H_n^*$:
\begin{equation}\label{eq:form2}
f_A=\sum_{i_1,\cdots, i_n}a_{i_1\cdots i_n}f_{i_1}^{(1)}\otimes\cdots\otimes
f_{i_n}^{(n)},
\end{equation}
where $f_i^{(k)}$ are standard $1$-forms on $H_k$ such that
$f_i^{(k)}(e_j^{(k)})=\delta_{ij}$ for
$i, j=1, \cdots, dim(V_i)$.

To examine the action of the SLOCC-symmetry on the hypermatrix representation
of $\rho$, we first give the following important lemma.

\begin{lemma} \label{T:mult} For fixed $A\in M_m$ and $C\in M_n$, let $\phi: M_{m, n}\longrightarrow M_{m, n}$
be the linear map defined by $B\mapsto ABC$. Then $det(\phi)=det(A)^ndet(C)^m$.
\end{lemma}
\noindent{ \bf Proof:} Choose a basis $B_1, \cdots, B_{mn}$ in $M_{m, n}$. Recall the
vector realignment $v(B)\equiv vec(B)$ \cite{HJ}. For any rectangular matrix $B=(b_{ij})\in M_{m,n}$, the
column vector realignment is given by $v(B)=[b_{11}, \ldots, b_{m1}, \cdots, b_{1n}, \ldots, b_{mn}]^t$.
The vector realignment satisfies the following fundamental property \cite{HJ}:
\begin{equation}
v(ABC)=(C^t\otimes A)v(B),
\end{equation}
where $C^t\otimes A$ is the Kronecker tensor product. It is also clear that
$v(B_1), \cdots, v(B_{mn})$ form a basis of $\mathbb C^{mn}$. Therefore the map $\phi$
induces an isomorphic linear operator on $M_{mn}\simeq \mathbb C^{mn}$ by
$\phi(v(B))=v(ABC)$ as $v(\,\cdot\, )$ is apparently linear. We will denote the
isomorphic map by the same symbol $\phi$. It then follows that
\begin{align*}
\phi[v(B_1), \cdots, v(B_{mn})]&=[v(AB_1C), \cdots, v(AB_{mn}C)]\\
&=(C^t\otimes A)[v(B_1), \cdots, v(B_{mn})],
\end{align*}
which means that $C^t\otimes A$ is the matrix of the linear operator $\phi\in End(\mathbb C^{mn})$.
Hence $det(\phi)=det(C^t\otimes A)=det(C)^mdet(A)^n$.
\qed

Consider a SLOCC symmetry $g=A_1\otimes\cdots\otimes A_n\in G=\mathrm{SL}(d_1)\otimes \cdots\otimes \mathrm{SL}(d_n)$
on the state space $H=H_1\otimes\cdots\otimes H_n$. This action is given by
$\rho\mapsto g\rho g^{\dagger}=(A_1\otimes\cdots\otimes A_n)\rho(A_1^{\dagger}
\otimes\cdots\otimes A_n^{\dagger})$. Let $B_k$ be the matrix of the action of $A_k$ on the orthonormal basis $\{\sigma_i^{(k)}\}$,
\begin{equation}\label{action-A}
A_k\sigma_i^{(k)}A_k^{\dagger}=\sum_{j=1}^{d_k^2}B_k^{ij}\sigma_i^{(k)},
\end{equation}
then the hypermatrix of $g\rho g^{\dagger}$ under the action of $g\in G$ is seen to be
\begin{equation}\label{action-g}
[g\rho g^{\dagger}]=(B_1*_1\cdots *_{n-1}B_n*_n)[\rho].
\end{equation}
One usually writes it as $(B_1\otimes\cdots\otimes B_n)[\rho]$ when
the hypermatrix $[\rho]$ is identified with its associated tensor via (\ref{eq:form2}).
This shows that the action of the SLOCC-symmetry on the
mixed state $\rho$ is represented by similar hypermatrix multiplications when one introduces the
matrix $B_k$ through
(\ref{action-A}) for each individual action of $A_k$ of the local SLOCC-symmetry $g=A_1\otimes\cdots\otimes A_n$.
The great advantage of (\ref{action-g}) is the similarity with the basis transition (\ref{action-P}).
The following result is immediate from Lemma \ref{T:mult} and the action (\ref{action-g}).

\begin{theorem}\label{T:SL} The action of $\mathrm{SLOCC}$-group $\mathrm{SL}(d_1)\otimes\cdots\otimes \mathrm{SL}(d_n)$ on $\rho$ induces an action of
$\mathrm{SL}(d_1^2)\times\cdots \times\mathrm{SL}(d_n^2)$ on its hypermatrix representation $[\rho]$.
\end{theorem}
We comment that a similar result also holds for a larger symmetry group $\widehat{SG}=\{g=A_1\otimes\cdots \otimes A_n|
det(g)=1, A_k\in \mathrm{GL}(d_k)\}$ when all $d_i$ are equal.

Theorem \ref{T:SL} translates the problem of SLOCC-invariants of $\rho$ into a specific problem
of invariants of its hypermatrix representation $[\rho]$ under the
group $\mathrm{SL}(d_1^2)\times\cdots \times\mathrm{SL}(d_n^2)$. This is exactly the classical problem
studied long ago by Cayley.

Cayley \cite{Ca} developed the theory of hyperdeterminant
as the theory of homogeneous polynomials in matrix elements
$A_{i_1\cdots i_n}$. A hyperdeterminant is a special
homogeneous polynomial invariant under the action of
$\mathrm{SL}(n_1)\times \cdots \times \mathrm{SL}(n_1))$.
Modern account of Cayley's theory
is the beautiful monograph \cite{gel}.  Cayley studied two types
of hyperdeterminants, and each is used in our current work.
The Cayley's (second) hyperdeterminant Det($A$) is defined as the
resultant of the multilinear form $f_A$, that is,
Det($A$) is certain
polynomial in components of the tensor $f_A$ which is zero if and only if the map $f_A$
has a non-trivial point where all partial derivatives with respect to the components of its vector arguments vanish (a non-trivial point means that none of the vector arguments are zero). For example, the resultant
of $a_1x_1^2+a_2x_1x_2+a_2x_2^2=0$ is the polynomial $a_2^2-4a_1a_2$.

It is known that the
hyperdeterminant Det($A$) exists for a given format and is unique up to a
scalar factor provided that any factor in the format is
less than or equal to the sum of the other factors in the format.
Cayley's two hyperdeterminants are all invariant
under the group $\mathrm{SL}(f_1) \otimes \cdots \otimes \mathrm{SL}(f_n)$.
In fact the hyperdeterminant $\mathrm{Det}$ satisfies the following multiplicative property.
Suppose $A$ is a hypermatrix of format $f_1\times\cdots\times f_n$
and $B$ is any $f_i\times f_i$ matrix, then
\begin{equation}\label{eq:det-action}
\mathrm{Det}(B*_iA) = \mathrm{Det}(A) \det(B)^{N/f_i},
\end{equation}
where $N$ is the degree of $\mathrm{Det}(A)$. Therefore we have
the following result.

\begin{theorem} The hyperdeterminant of the matrix of $\rho$ is a
relative invariant under
the action of $\mathrm{GL}(d_1) \otimes \cdots \otimes \mathrm{GL}(d_n)$,
thus invariant under the SLOCC symmetry group $G=\mathrm{SL}(d_1) \otimes \cdots
\otimes \mathrm{SL}(d_n)$.
\end{theorem}

For an even dimensional matrix $A=(A_{i_1\cdots i_{2n}})$, where
$1\leq i_1,\ldots, i_{2n}\leq N$, the
Cayley's first hyperdeterminant $hdet(A)$ is the following
polynomial
\begin{equation}\label{hdet}
hdet(A)=\frac1{N!}\sum_{\sigma_1, \cdots, \sigma_{2n}\in S_N}sgn(\sigma)
\prod_{i=1}^{N}A_{\sigma_1(i), \ldots, \sigma_{2n}(i)},
\end{equation}
where $sgn(\sigma)=\prod_{i=1}^n sgn(\sigma_i)$.
The hyperdeterminant $hdet$ reduces to the usual determinant when $A$ is a square matrix.

Cayley's second hyperdeterminant \cite{Ca}
of the format $2\times
2\times 2$ for the hypermatrix $A$ with components
$a_{ijk}$, $i,j,k \in \{0, 1\}$, is given by
$$
\begin{array}{rl}
\mathrm{Det}(A)=& a_{000}^2a_{111}^2 + a_{001}^2a_{110}^2 +
a_{010}^2a_{101}^2 + a_{100}^2a_{011}^2\\[2mm]
&-2a_{000}a_{001}a_{110}a_{111}-2a_{000}a_{010}a_{101}a_{111}\\[2mm]
&-2a_{000}a_{011}a_{100}a_{111} - 2a_{001}a_{010}a_{101}a_{110}\\[2mm]
&-2a_{001}a_{011}a_{110}a_{100}- 2a_{010}a_{011}a_{101}a_{100}\\[2mm]
&+4a_{000}a_{011}a_{101}a_{110}+4a_{001}a_{010}a_{100}a_{111}.
\end{array}
$$
This hyperdeterminant can be written in a more compact form by using
the Einstein convention and the Levi-Civita symbol
$\varepsilon^{ij}$, with $\varepsilon^{00} =\varepsilon^{11} = 0$,
$\varepsilon^{01} = -\varepsilon^{10} = 1$; and $b_{kn} =
(1/2)\varepsilon^{il}\varepsilon^{jm}a_{ijk}a_{lmn}$, $ \mathrm{Det}(A)
=(1/2)\varepsilon^{il}\varepsilon^{jm}b_{ij}b_{lm}$. The
four-dimensional hyperdeterminant of the format $2\times 2\times 2
\times 2$ has been given in Ref. \cite{Luq}.

In order to generate more invariants
out of the hyperdeterminants,
observe that the tensor product  $I_{\mathbf n}=I_{d_1}\otimes\cdots \otimes I_{d_n}$
is a special hypermatrix of format $d_1^2\times\cdots\times d_d^2$
provided one views it as a tensor.
Moreover for any $A=A_1\otimes\cdots\otimes A_n\in \hat{G}$, one has that
\begin{equation}
(A_1*_1\cdots *_{n-1}A_n*_n)I_{\mathbf n}=A_1\otimes\cdots\otimes A_n
\end{equation}
which is also an element of $Mat(d_1^2, \cdots, d_n^2)$.

\begin{theorem} (1) The hyper-characteristic polynomial
$\mathrm{Det}(\lambda I_{\mathbf{n}}-\rho)$ $=\mathrm{Det} (\lambda I_{\mathbf{n}}-[\rho])$ of the multipartite
mixed state $\rho$ with
$[\rho]=(A_{i_1\cdots i_n})\in Mat(d_1^2, \cdots, d_n^2)$, is a SLOCC-invariant polynomial in $\lambda$. The coefficients of
the characteristic polynomial are all SLOCC-invariants. In particular, the hyper-trace
$Tr(A)$ and the hyperdeterminant $\mathrm{Det}(A)$ of $A$ are two distinguished polynomial SLOCC-invariants.
(2) The same results hold for even-partite quantum states when $\mathrm{Det}$ is replaced with $hdet$.
\end{theorem}
\noindent{\bf Proof:} Both hyperdeterminants are proved similarly,
so we only consider $\mathrm{Det}$. By previous discussion either the transition matrix
or the action of the SLOCC-symmetry is represented by
the hyper-matrix multiplication, see (\ref{action-P}) and (\ref{action-g}):
\begin{align*}
[\rho]_{\sigma'}&=(P_1*_1\cdots *_{n-1}*P_n*_n)[\rho]_{\sigma},\\
[g\rho g^{\dagger}]&=(B_1*_1\cdots *_{n-1}*B_n*_n)[\rho].
\end{align*}
Therefore the hyper-characteristic polynomial of
$g\rho g^{\dagger}$ is
\begin{align*}
&\mathrm{Det}\left((B_1*_1\cdots *_{n-1}B_n*_n)(\lambda I_{\mathbf n}-[\rho])\right)\\
&=det(B_1)^{N/d_1}\cdots det(B_1)^{N/d_1}\mathrm{Det}(\lambda I_{\mathrm n}-\rho)\\
&=\mathrm{Det}(\lambda I_{\mathrm n}-\rho).
\end{align*}
This means that $f(\lambda)=\mathrm{Det}(\lambda I_{\mathbf n}-\rho)$ is a SLOCC-invariant polynomial and every $\lambda$-coefficients
of $f(\lambda)$ are SLOCC-invariants. Note that the above calculation also proves the
independence from base change.
\qed

We emphasize that in general the state $I_{\mathbf n}$ is not represented by a tensor product of the identity
operators except for the special bipartite case. This means that although in formality our result
resembles the usual theory of characteristic polynomial, but the result contains substantially more
information in general and constitutes a theoretical method to generate non-trivial invariants for the
mixed states.

As an application of our general method, we give a few examples.

Example 1. For two-qubit case, $A=(a_{ij})\in M_4$. The followings are SLOCC-invariants.
\begin{align*}
&Tr(A);\qquad \det(A);\\
&m_{12}+m_{13}+m_{14}+m_{23}+m_{24}+m_{34};\\
&m_{234}+m_{134}+m_{124}+m_{234},
\end{align*}
where $m_{ij}$ etc are the principal minors, i.e. $m_{12}$
is the $(1, 2)$-principal minor of $A$.

Example 2. When $\rho$ is a mixed state on $H=\bigotimes_1^{2n}{\mathbb C}^d$, we will get $d^2$ SLOCC invariants
from $hdet(\lambda I_{\mathbf n}-\rho)$. The four nontrivial coefficients of the following
polynomial provides SLOCC-invariants for the mixed $4$-qubit state.
\begin{align*}
&hdet(A-\lambda I)=\\
&\frac1{24}\sum_{\sigma\in S_4^3}sgn(\sigma)\prod_{i=1}^4(A_{\sigma_1(i)\sigma_2(i)\sigma_3(i)\sigma_4(i)}-\lambda \delta_{\sigma_1(i)\sigma_2(i)}\delta_{\sigma_3(i)\sigma_4(i)}).
\end{align*}

\section{Conclusion and Discussion}
We have introduced a general set of SLOCC-invariants for mixed multipartite states
by using hyperdeterminants of the matrix representation in terms of
orthonormal bases. We have shown that these invariants are independent of matrix
representations.

Our method is fundamentally different from previous
studies on LU-invariants of pure states, and enjoys
favorable property of independence from
base change. In fact, our method
shows that as long as the transition matrix belongs
to $\mathrm{SL}(n)$ then the hypermatrix
representation of the mixed state $\rho$ can deliver
the same result. Even in the case of pure quantum states,
our method is also different.
Suppose that a pure state is given by $|\phi\rangle=\sum_{I}a_I|I\rangle$,
and assume that the transition matrix from $|I\rangle\langle J|$ to $\lambda$-basis
is $(p_{IJ,\sigma})$, then the hypermatrix of $\rho=|\phi\rangle\langle\phi|$ is
$(\sum_{IJ}a_Ia^*_Jp_{IJ,\sigma})$, whose hyperdeterminant
is quite different from that of $(a_{I})$, even in the format.

The method of hyperdeterminants can be used for classification of multipartite
quantum states besides theoretical classification of SLOCC symmetry.
Since the invariants involve with
large amount of computation, in practice they should be useful with the help of computer
codes. In our opinion one of the useful messages is that many LU-invariants obtained from
the hypermatrix representation are actually SLOCC-invariants.

\medskip

\centerline{\bf Acknowledgments} NJ gratefully acknowledges the
partial support of Simons Foundation, Humbolt Foundation, 
NSF, NSFC 
and MPI for Mathematics in the Sciences in Leipzig during this work.


\begin{thebibliography}{99}

\bibitem{GW} Gour G and Wallach N R, 2013, Phys. Rev. Lett. {\bf 111}, 060502; 2010, J. Math Phys. {\bf 51}, 112201.

\bibitem{YQS} Yu N, Qiao Y and Sun X, Characterization of multipartite entanglement,
arXiv:1401.2627.

\bibitem{Laura} Albeverio S, Cattaneo L and Persio D L, 2007, Rep. Math. Phys. {\bf 60}, 167.

\bibitem{FJ} Fei S M and Jing N, 2005, Phys. Lett. A {\bf 342}, 77.

\bibitem{Grassl} Grassl M R, R\"otteler M and Beth T, 1998, Phys. Rev. A. {\bf 58}, 1833.

\bibitem{Rains} Rains E M, 2000, IEEE Trans. Inf. Theory. {\bf 46}, 54.

\bibitem{makhlin} Makhlin Y, 2002, Quant. Info. Proc. {\bf 1}, 243.

\bibitem{linden} Linden N, Popescu S and Sudbery A, 1999, Phys. Rev. Lett. {\bf 83}, 243.

\bibitem{SCFW} Albeverio S, Cattaneo L, Fei S M and Wang X H, 2005, Int. J. Quant. Inform. {\bf3}, 603.

\bibitem{SFG} Albeverio S, Fei S M and Goswami D, 2005, Phys. Lett. A. {\bf 340}, 37.

\bibitem{SFY} Albeverio S, Fei S M, Parashar P and Yang W L, 2003, Phys. Rev. A. {\bf 68}, 010303.

\bibitem{SFW} Sun B Z, Fei S M, Li-Jost X Q and Wang Z X, 2006, J. Phys. A. {\bf 39}, 43.

\bibitem{mqubit} Kraus B, 2010, Phys. Rev. Lett. {\bf 104}, 020504; 2010, Phys. Rev. A {\bf 82}, 032121.

\bibitem{B. Liu} Liu B, Li J L, Li X and Qiao C F, 2012, Phys. Rev. Lett. {\bf 108}, 050501.

\bibitem{Ca} Cayley A, 1845, Cambridge Math. J. {\bf 4}, 193.

\bibitem{gel} Gelfand I M, Kapranov M M and Zelevinsky A V, 1994, Discriminants, Resaultants, and Multidimensional Determinants.
Birkhaeuser, Boston.

\bibitem{ass} Albeverio S and Fei S M, 2001, J. Opt. B: Quantum Semiclass. Opt. {\bf 3}, 223.

\bibitem{Hill} Hill S and Wootters W K, 1997, Phys. Rev. Lett. {\bf 78}, 5022.

\bibitem{rungta} Rungta P, Buzek V, Caves C M,  Hillery M and Milburn G J, 2001, Phys. Rev. A. {\bf 64}, 042315.

\bibitem{uhlm} Uhlmann A, 2000, Phys. Rev. A. {\bf 62}, 032307.

\bibitem{wott} Wootters W K, 1998, Phys. Rev. Lett. {\bf 80}, 2245.

\bibitem{coff} Coffman V, Kundu J and Wootters W K, 2000, Phys. Rev. A. {\bf 61}, 052306.

\bibitem{Luq} Luque J G and Thibon J Y, 2003, Phys. Rev. A. {\bf 67}, 042303.

\bibitem{miy} Miyake A, 2003, Phys. Rev. A. {\bf 67}, 012108.

\bibitem{vie} Viehmann O, Eltschka C and Siewert J, 2011, Phys. Rev. A {\bf 83}, 052330.

\bibitem{ZJLZF} Zhang T G, Jing N, Li-Jost X Q, Zhao M J and Fei S M, 2013, Euro. Phys. J. D {\bf 67}, 175.

\bibitem{HJ} Horn R A and Johnson C R, 1994, Topics in matrix analysis. Cambridge University Press, Cambridge.

\end{thebibliography}
\end{document}